\documentclass[twocolumn, switch]{article} 
\usepackage[colorlinks = true,
            linkcolor = purple,
            urlcolor  = blue,
            citecolor = cyan,
            anchorcolor = black]{hyperref}	
\usepackage{preprint}

\usepackage{amsmath, amsthm, amssymb, amsfonts}

\usepackage[numbers,square]{natbib}
\usepackage[utf8]{inputenc}	
\usepackage[T1]{fontenc}	
\usepackage{xcolor}		
\usepackage{booktabs} 		
\usepackage{nicefrac}		
\usepackage{microtype}		
\usepackage{lineno}		
\usepackage{float}			

\usepackage{lipsum}		
\usepackage{newfloat}
\DeclareFloatingEnvironment[name={Supplementary Figure}]{suppfigure}
\usepackage{sidecap}
\sidecaptionvpos{figure}{c}

\usepackage{titlesec}
\titlespacing\section{0pt}{12pt plus 3pt minus 3pt}{1pt plus 1pt minus 1pt}
\titlespacing\subsection{0pt}{10pt plus 3pt minus 3pt}{1pt plus 1pt minus 1pt}
\titlespacing\subsubsection{0pt}{8pt plus 3pt minus 3pt}{1pt plus 1pt minus 1pt}

\title{Detection of electron spin resonance down to 10 K using localized  spoof surface plasmon}

\usepackage{xwatermark}
\newwatermark[firstpage,color=gray!90,angle=0,scale=0.28, xpos=0in,ypos=-5in]{*correspondence: \texttt{chiranjib@iiserkol.ac.in}}

\usepackage{authblk}

\author[ ]{Subhadip Roy}
\author[ ]{Anuvab Nandi}
\author[ ]{Pronoy Das}
\author[*]{Chiranjib Mitra}
\affil[ ]{Department of Physical Sciences, Indian Institute of Science Education and Research Kolkata, India.}

\begin{document}

\twocolumn[ 
  \begin{@twocolumnfalse} 
  
\maketitle 

\begin{abstract}
In this study, novel use of the electromagnetic field profile of a localized spoof surface plasmonic mode to detect electron spin resonance is being reported. The mode is supported on a resonator with a complementary metallic spiral structure, etched on the ground plane of a microstrip line having a characteristic impedance of 50 $\Omega$. The change in characteristics of the mode of interest with lowering of temperature has been observed and analyzed. Electron spin resonance spectra of a standard paramagnetic sample, 2,2-diphenyl-1-picrylhydrazyl, are recorded using this resonator down to 10 K. Potential application of the mode in the detection of microwave Rashba field-driven electron spin resonance has been discussed.  
\end{abstract}
\vspace{0.35cm}

  \end{@twocolumnfalse} 
] 



\section{Introduction}
\label{intro}
Surface plasmons are electromagnetic waves which exist on metal and dielectric interface at optical frequencies \cite{sp},\cite{sp1},\cite{sp3}. They can either propagate at the interface as surface plasmon polaritons or can be resonant in the form of localized surface plasmons (LSPs)  \cite{sp_types},\cite{types}. However, at low-frequency range such as at microwave and terahertz frequencies, metals behave as nearly perfect electric conductors which prevent the  excitation of these surface modes \cite{surface_plasmon}. Patterned metallic surfaces can support surface waves at low frequencies, which have features similar to original surface plasmons, and are known as spoof surface plasmons \cite{actual_mode},\cite{actual_mode1},\cite{spoof}. Similar to the original LSPs, the localized spoof surface plasmons (LSSPs) show strong field confinement \cite{confine},\cite{confine1}. This property of LSSPs has been utilized in this work for the detection of electron spin resonance (ESR) and related measurements. Previously, LSSP modes have been used for sensing purposes only \cite{prev_app},\cite{lssp_ms}.\\
ESR occurs when a resonant microwave radiation cause transition of electrons between spin levels in the presence of an external Zeeman magnetic field in paramagnetic systems \cite{ESR}. The sample under study is placed in the region of a uniform microwave magnetic field, orthogonal to the Zeeman field to cause ESR transitions \cite{perp}. So far, cavity resonators \cite{cavity}, lumped resonators \cite{lgr} and planar resonators \cite{planar} have been used for microwave magnetic field generation at the position of the sample.\\
LSSP modes can be generated on a complementary metallic spiral structure (CMSS) \cite{CMSS},\cite{CMSS1}. An LSSP mode supported on a resonator, consisting of a CMSS excited by a microstrip line \cite{lssp_ms} has been used in this work for detection of ESR. The LSSP resonator has been simulated, and the fabricated resonator is used to record ESR spectra of 2,2-diphenyl-1-picrylhydrazyl (DPPH) down to 10 K. In section \ref{sec2}, detailed description of the LSSP resonator has been provided. Elaborate description of low-temperature setup for recording ESR spectra is done in section \ref{sec3}, followed by section \ref{sec4} containing the results and related discussion. The study is concluded in section \ref{sec5}, with a discussion on the potential application of the designed resonator in the detection of ESR transitions caused by a microwave Rashba field in low dimensional systems.
 
\section{The Localized Spoof Surface Plasmonic Resonator}
\label{sec2}
\subsection{Geometry}
\label{geometry}
\begin{figure}[H]
    \centering
    \includegraphics[scale=0.3]{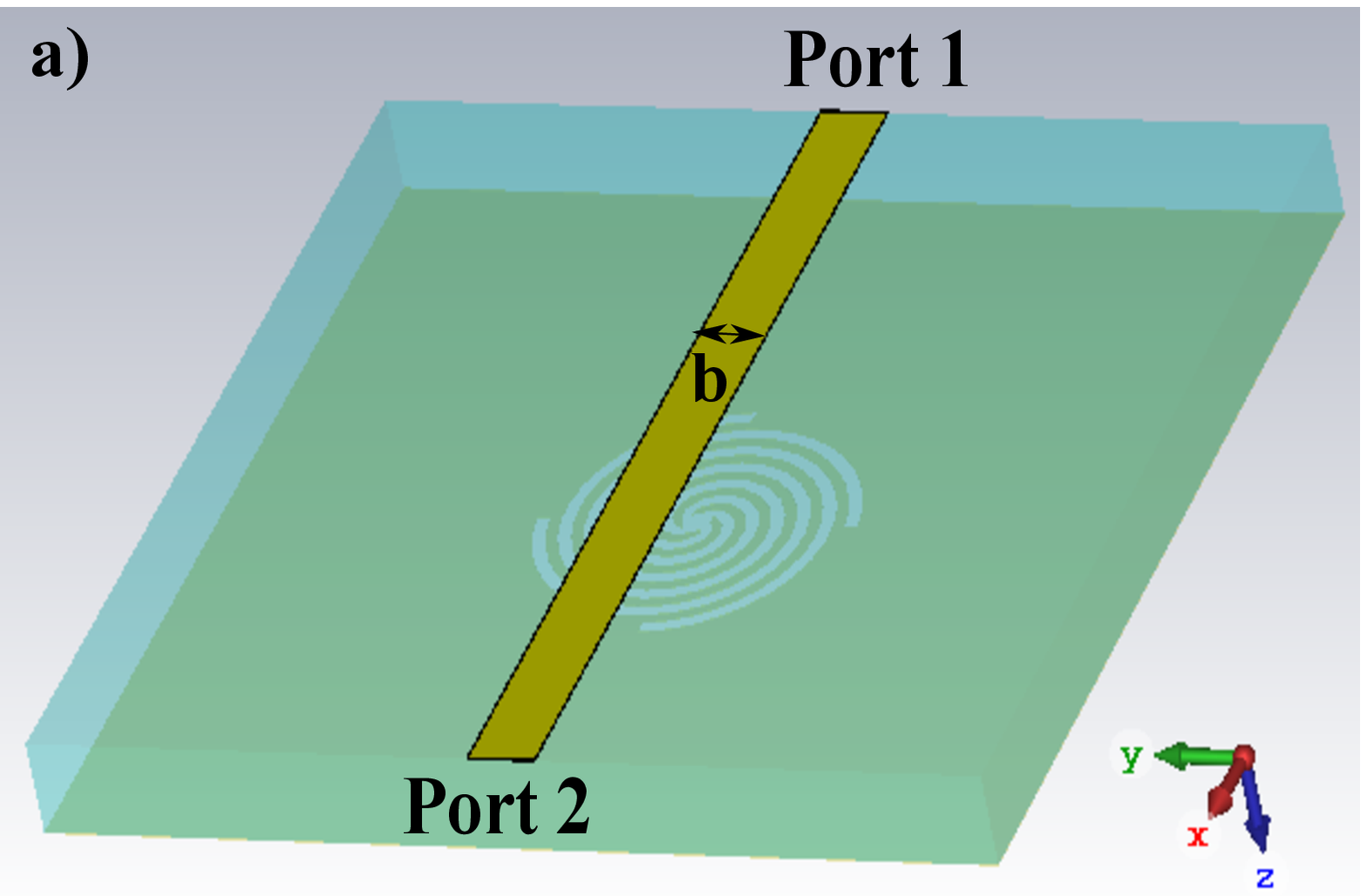}
    \includegraphics[scale=0.06]{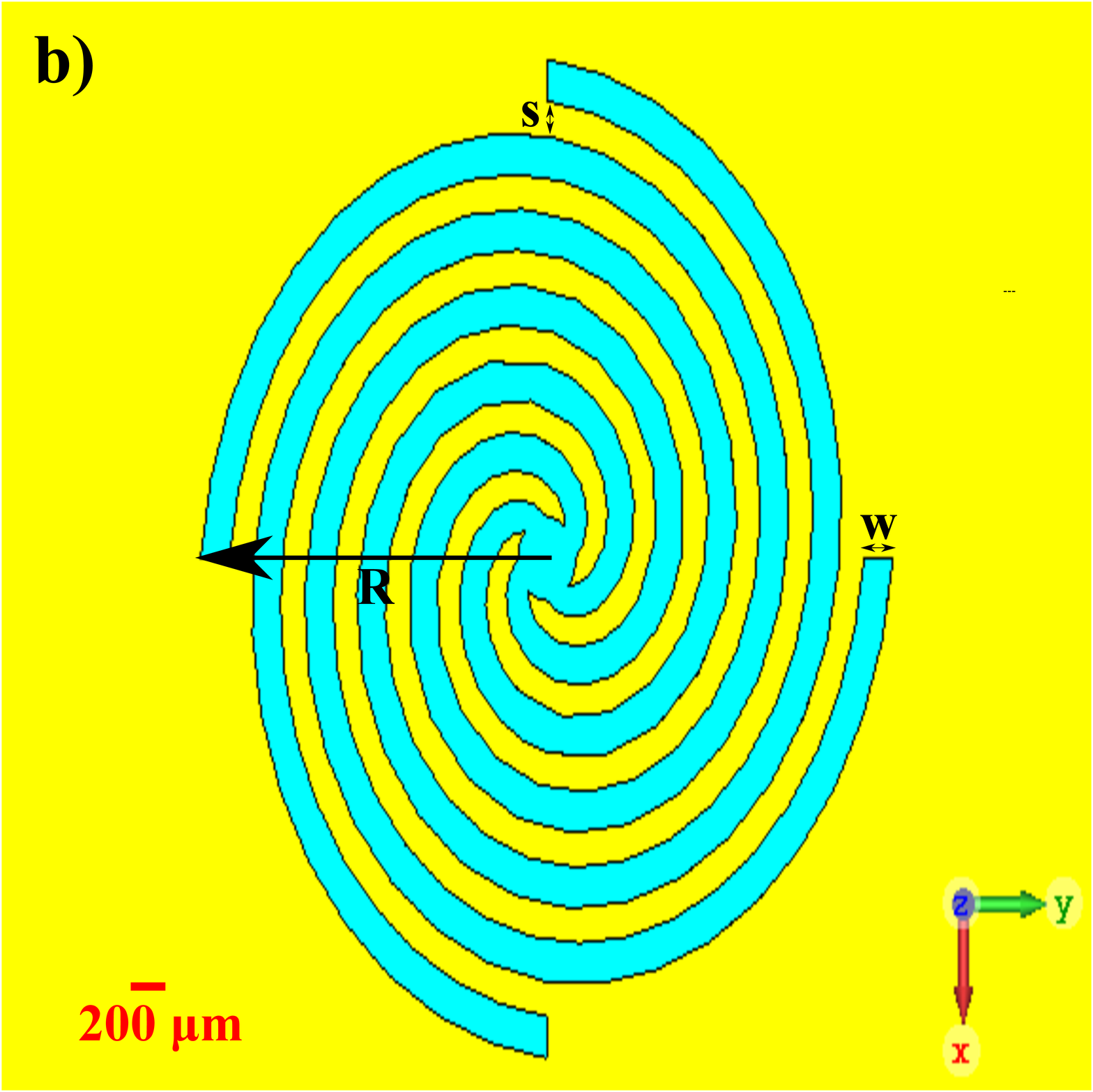}
    \caption{(a) The LSSPR along with excitation ports. (b) The CMSS located on the ground plane with dimensions marked.}
    \label{CMSS}
\end{figure}

The Localized Spoof Surface Plasmonic Resonator (LSSPR) comprises a CMSS etched on the ground plane of a microstrip transmission line along with conducting trace on the other side which excites the LSSP modes. Four spiral air slots of width \textbf{w} = 0.24 mm and having a separation \textbf{s} = 0.2 mm with respect to adjacent slot wraps 1.5 times to form the CMSS structure with outer radius \textbf{R} = 2.88 mm. The signal trace on the opposite face has a width \textbf{b} = 1.2 mm, which corresponds to 50 $\Omega$ characteristic impedance. Figures \ref{CMSS} (a) \& (b) illustrate the LSSPR and the CMSS, respectively.
\subsection{Simulation \& Fabrication}
\begin{figure}[H]
    \centering
    \includegraphics[scale=0.08]{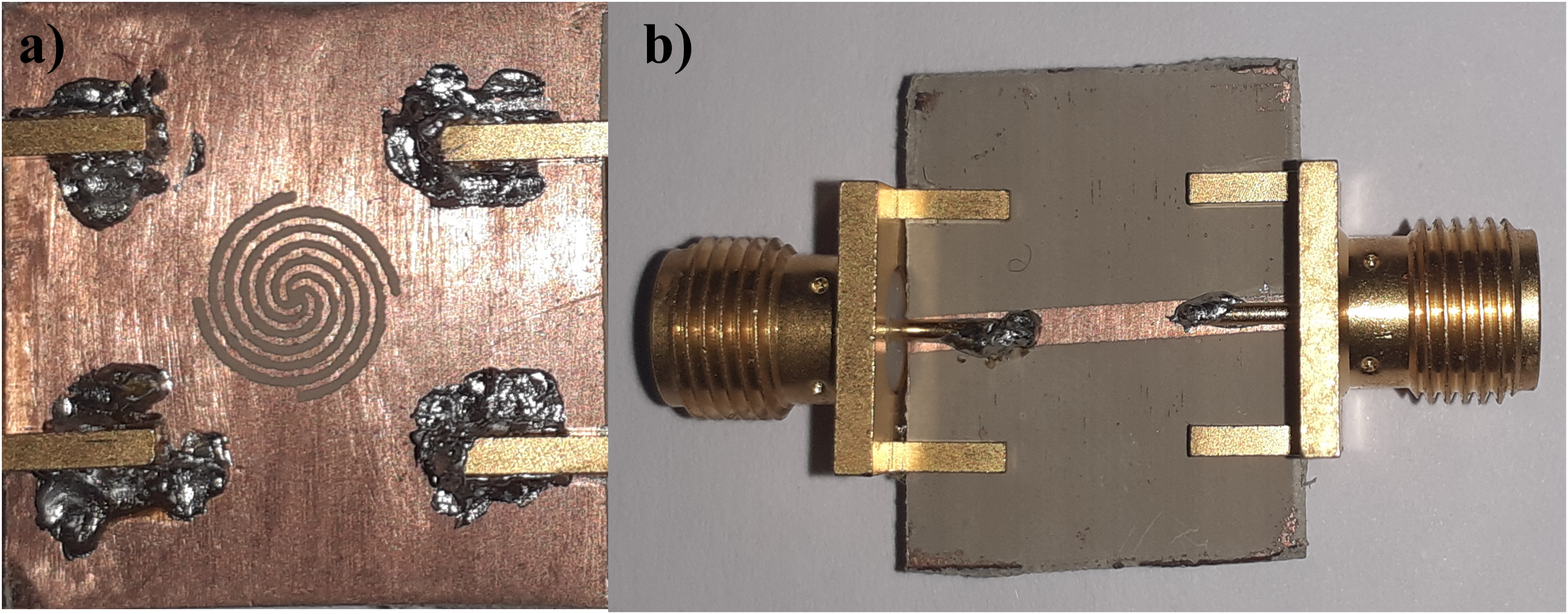}
    \caption{(a) The CMSS structure after fabrication. (b) The signal trace with SMA connectors soldered on it.}
    \label{fab}
\end{figure}

Simulation of the LSSPR having dimensions indicated in sub-section \ref{geometry} is carried out in CST Microwave Studio (CST MWS) software. The microwave laminate used in the process is AD1000 \cite{Rogers} (Rogers Corporation,  USA). The laminate has a dielectric constant of 10.7 for a thickness of 1.5 mm and a dissipation factor of 0.0023 defined at 10 GHz. A 17.5 $\mu$m thick copper layer is present on both sides of the dielectric. In the simulation setup, the electrical conductivity of copper is taken as 5.96$\times$ 10$^7$ S/m, which is the pre-defined value available in CST MWS. Frequency-domain solver with open (add space) boundary condition on the ground plane side and open boundary condition on all other sides of the structure has been used while running the simulation.\\
The fabrication of the structure is carried out using a rapid prototyping process employing optical lithography \cite{litho} followed by wet etching \cite{wet}. SMA connectors are soldered at either end of the conducting trace for the propagation of  microwave signal. The fabricated LSSPR with the CMSS on the ground plane and the signal trace are shown in figures \ref{fab} (a) \& (b) respectively.
\subsection{Characterization}
\begin{figure}[H]
    \centering
    \includegraphics[scale=0.3]{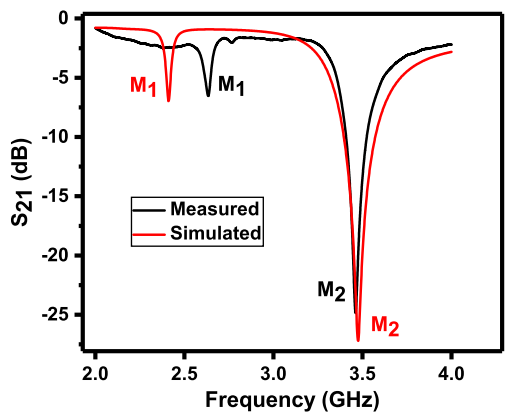}
    \caption{Comparison of the simulated (red) \& measured (black) transmission spectra of the LSSPR in 2-4 GHz range.}
    \label{M1M2}
\end{figure}

The LSSPR is characterized by measuring its transmission spectrum and comparing it with the simulated result. The measurement is carried out using a Vector Network Analyzer (VNA) (ZVA 24, Rohde \& Schwarz). The VNA is calibrated using through, open, short and match (TOSM) standards before performing the measurements. Figure \ref{M1M2} compares the measured response (black) with the simulated one (red). The deviation between the two responses may be attributed to the tolerance of the fabrication process. In the 2-4 GHz frequency range, the transmission spectrum shows two dips marked as $\textbf{M$_1$}$ \& $\textbf{M$_2$}$ which are the two fundamental LSSP resonance modes. $\textbf{M$_1$}$ is the magnetic LSSP mode and $\textbf{M$_2$}$ is the electrical LSSP mode \cite{lssp_ms},\cite{CMSS}.\\

\section{Application of the LSSPR in detection of electron spin resonance down to 10 K} 
\label{sec3}
\begin{figure}[H]
    \centering
    \includegraphics[scale=0.2]{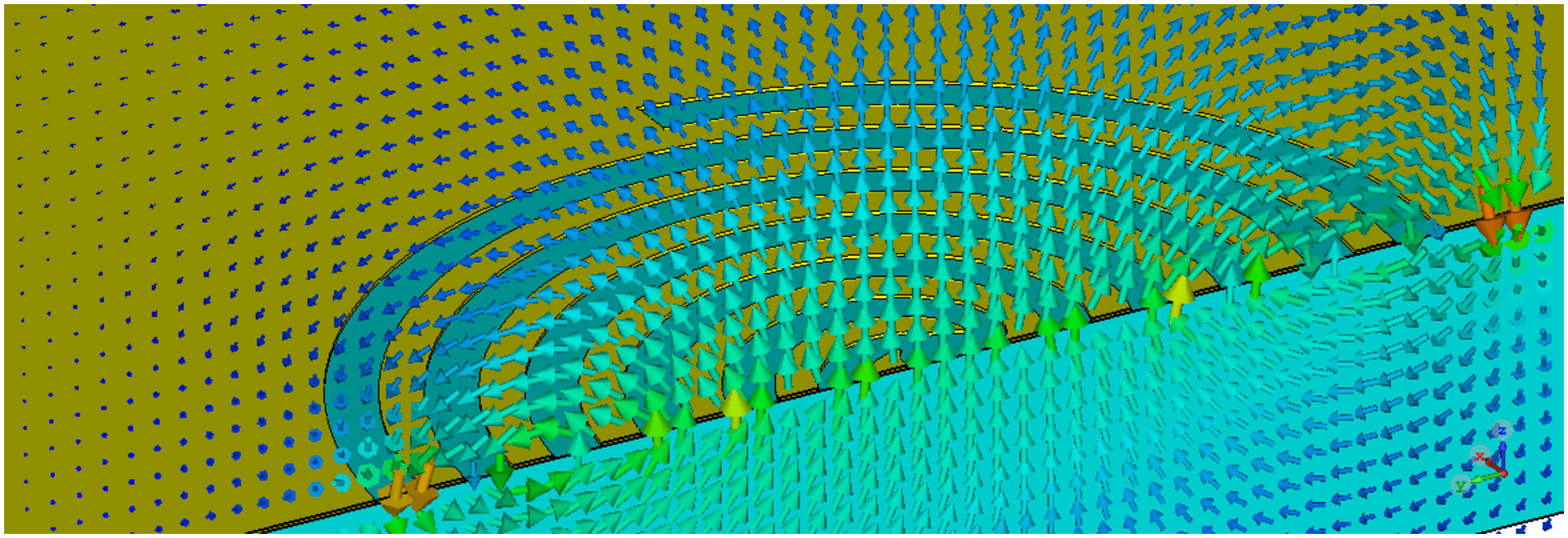}
    \caption{Cross-sectional view of the simulated microwave magnetic field distribution in the $yz$ plane.}
    \label{magfielddirection}
\end{figure}

\begin{figure}[H]
    \centering
    \includegraphics[scale=0.4]{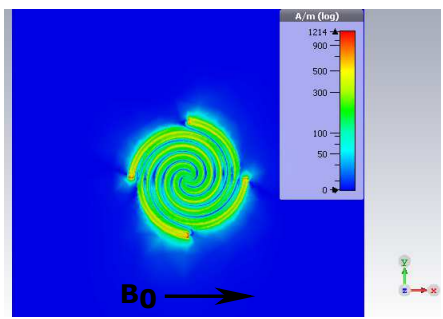}
    \caption{Distribution of |$\textbf{B$_\perp$}$| just above the CMSS structure. The direction of $\textbf{B$_0$}$ along $x$-axis is indicated.}
    \label{B_perp}
\end{figure}

\begin{figure}[H]
    \centering
    \includegraphics[scale=0.4]{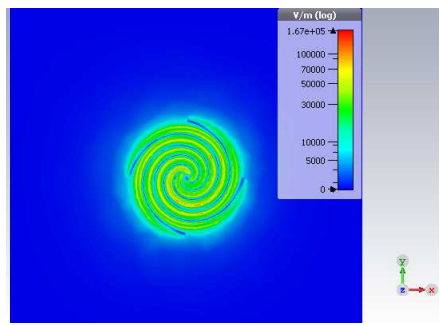}
    \caption{Distribution of the electric field just above the CMSS structure.}
    \label{E_field}
\end{figure}

\begin{figure}[H]
    \centering
    \includegraphics[scale=0.08]{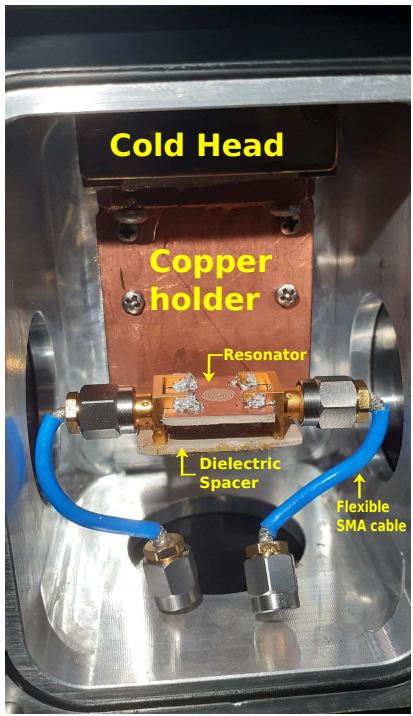}
    \caption{The mounted empty LSSPR inside the cryostat.}
    \label{low_temp}
\end{figure}

The LSSP mode $\textbf{M$_2$}$ resonating at \textbf{3.46} GHz (measured) as shown in figure \ref{M1M2} is chosen for the application in the detection of electron spin resonance.  A uniform magnetic field is obtained in the central part of the resonator as shown in figure \ref{magfielddirection}.
The Zeeman field $\mathbf{B_0}$ is applied along the $x$-direction as shown in figure \ref{B_perp}. Hence the magnitude of the component of the microwave magnetic field perpendicular to $\mathbf{B_0}$ is given by $|\mathbf{B_\perp}| = \sqrt{{|\mathbf{B_{1z}}|}^2+|\mathbf{B_{1y}|}}^2$. Figures \ref{B_perp} and \ref{E_field} show the distribution of $|\mathbf{B_\perp}|$ and the electric field just above the CMSS structure, respectively. 
Temperature variation of the $\textbf{M$_2$}$ mode's properties is investigated by mounting the LSSPR inside a closed-cycle  cryogen-free cryostat (\textbf{Optistat}Dry BLV, Oxford Instruments) on a custom designed copper holder attached to the cold head. The stability of temperature  during measurements is maintained by using a temperature controller (\textbf{Mercury}iTC, Oxford Instruments). The LSSPR is attached to a dielectric spacer for electrical insulation which in turn is stuck to the holder. Apiezon N grease and a cyanoacrylate adhesive are used to provide thermal contact and mechanical stability respectively. The arrangement is depicted in Figure \ref{low_temp}. Hand formable microwave cables (086-2SM+, Mini-Circuits) connect the LSSPR to the VNA via hermetically sealed adapters (PE9184, Pasternack) fitted to the cryostat body. TOSM calibration is performed at room temperature before the commencement of measurement at low temperatures.

\begin{figure}[H]
    \centering
    \includegraphics[scale=0.5]{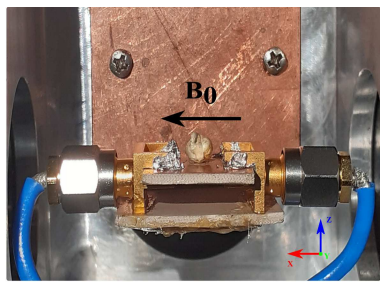}
    \caption{Location of the sample on the LSSPR}
    \label{sample}
\end{figure}

\begin{figure}[H]
    \centering
    \includegraphics[scale=0.028]{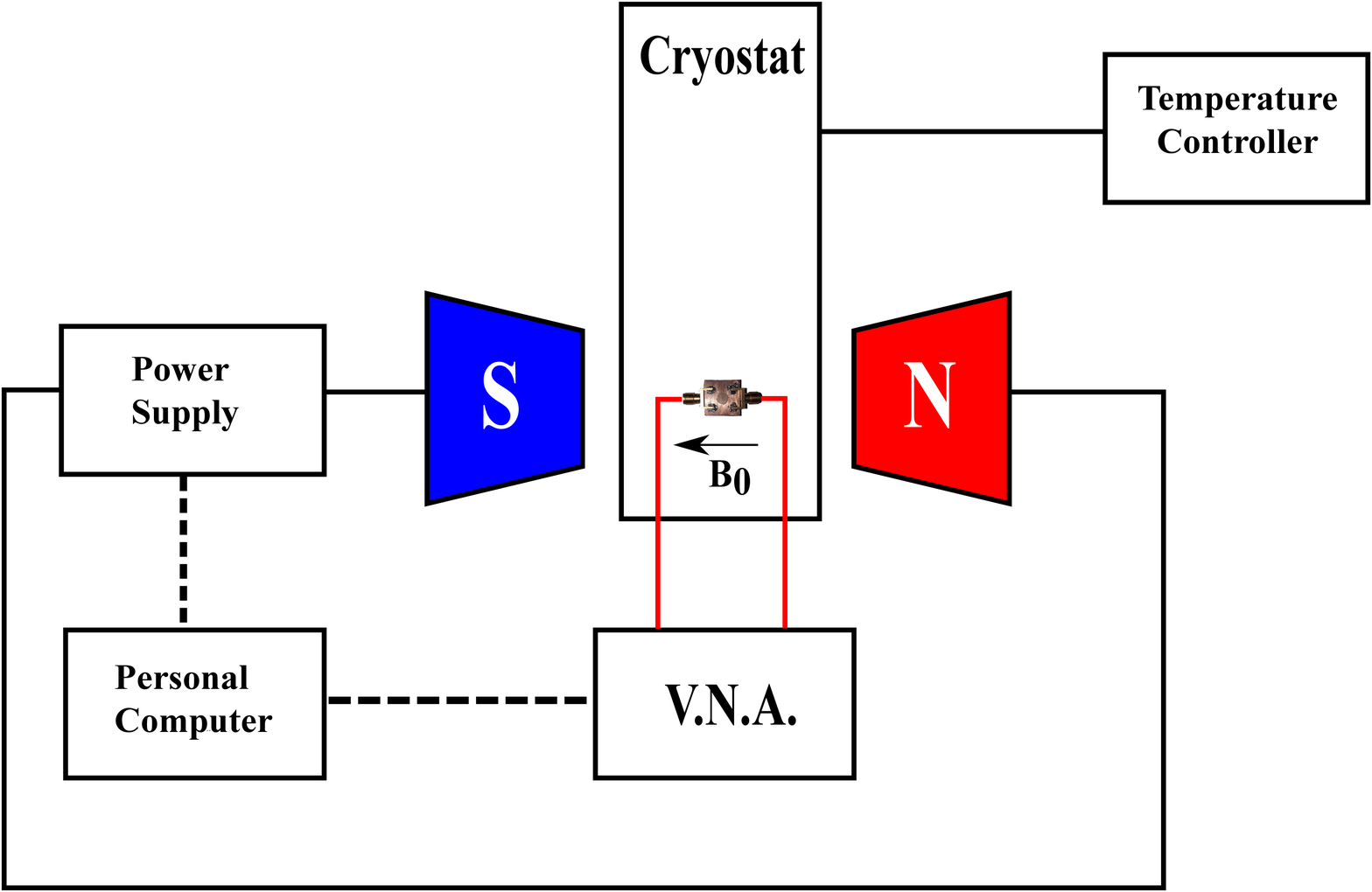}
    \caption{Schematic of the low temperature ESR setup}
    \label{esr_setup}
\end{figure}

Continuous wave electron spin resonance spectroscopy is performed on 4mg of DPPH sample at different temperature values in the 10 K to 295 K temperature range. The powder sample is wrapped in teflon tape and is affixed on the CMSS structure with the help of Apiezon N grease. It is placed on the central part
of the resonator as the microwave magnetic field is uniform there. Figure \ref{sample} shows the sample location on the LSSPR. 
The cryostat loaded with the sample is placed between the pole pieces of an electromagnet (3473-70, GMW). The electromagnet provides the external Zeeman field. The VNA connected to the LSSPR acts as the source and detector of microwaves. It is set to have a measurement bandwidth of 1 MHz with a frequency step size of 500 kHz. An averaging factor of 25 with 15 dBm port power is used for recording the ESR spectra. The programmable power supply (SGA60X83D, Sorensen) connected to the electromagnet, and the VNA are interfaced using a Python script. Temperature is manually set on the temperature controller before recording an ESR spectrum. The schematic of the low temperature ESR setup is shown in figure \ref{esr_setup}.
\section{Results \& Discussion}
\label{sec4}
\begin{figure}[H]
    \centering
    \includegraphics[scale=0.3]{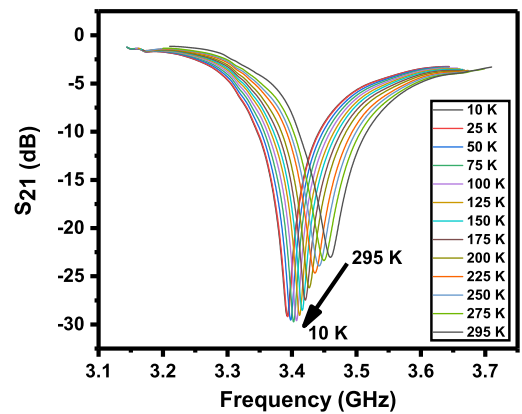}
    \caption{Temperature evolution of \textbf{M$_2$} mode for the empty LSSPR.}
    \label{M2_empty}
\end{figure}

\begin{figure}[H]
    \centering
    \includegraphics[scale=0.32]{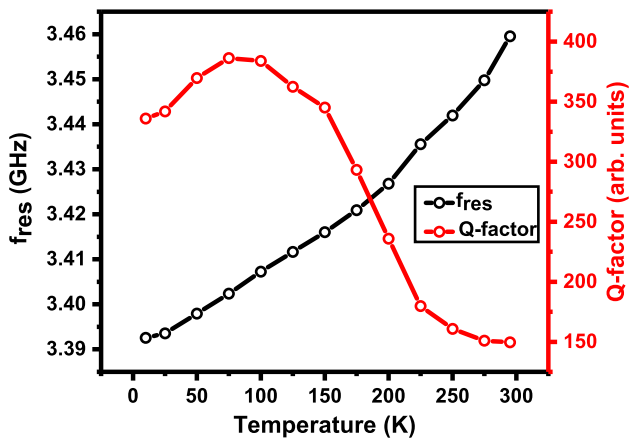}
    \caption{Change in  the resonant frequency,  \textbf{f\textsubscript{res}} (black) and Q-factor (red) with temperature for the empty LSSPR.}
    \label{temp_var}
\end{figure}

The temperature evolution of $\textbf{M$_2$}$ mode for the empty LSSPR is plotted in figure \ref{M2_empty}. The shift of the resonant frequency (\textbf{f\textsubscript{res}}) and the variation in the loaded quality factor (Q-factor) of the $\textbf{M$_2$}$ mode for the empty LSSPR with temperature are shown in figure \ref{temp_var}. The Q-factor is calculated using the relation, Q-factor = $\frac{\text{\textbf{f\textsubscript{res}}}}{\Delta f_{3dB}}$, where $\Delta f_{3dB}$  is the bandwidth at +3dB above the minimum forward transmission coefficient S$_{21}$ value which occurs at \textbf{f\textsubscript{res}} \cite{qfactor},\cite{qfactor1},\cite{qfactor2}.

\begin{figure}[H]
    \centering
    \includegraphics[scale=0.32]{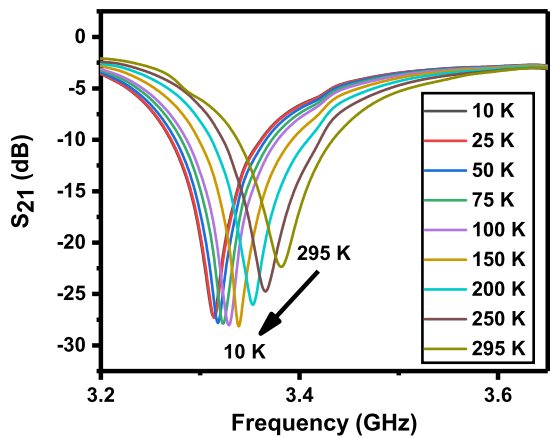}
    \caption{Temperature evolution of \textbf{M$_2$} mode for the LSSPR loaded with sample.}
    \label{M2_sample}
\end{figure}

\begin{figure}[H]
    \centering
    \includegraphics[scale=0.19
]{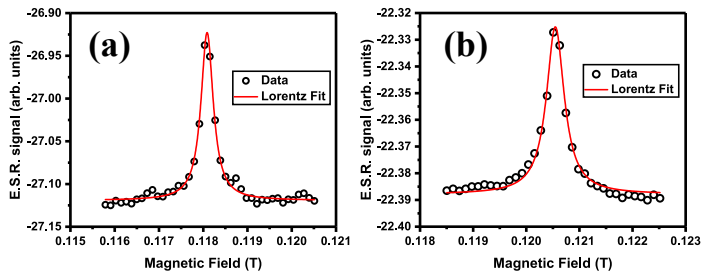}
    \caption{ESR signal fitted by a Lorentzian curve (red) recorded at (a) 10 K (b) 295 K (room temperature) }
    \label{ESR}
\end{figure}

\begin{figure}[H]
    \centering
    \includegraphics[scale=0.18]{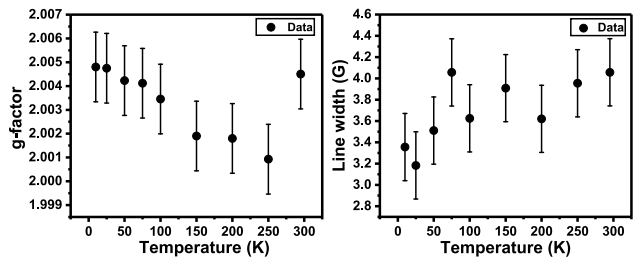}
    \caption{Measured (a) g-factor \& (b) ESR line width at different temperature values.}
    \label{result}
\end{figure}

A shift of 67 MHz is observed in \textbf{f\textsubscript{res}} at 10 K in comparison to its value at 295 K. Thermal contraction and change in dielectric properties of the microwave laminate \cite{Rogers} can qualitatively explain this shift \cite{shift}, \cite{shift1}, \cite{loss}. The value of the Q-factor reaches a maximum of 386 at 75 K after which it decreases slightly down to 10 K. The enhancement in Q-factor at low temperature can be attributed to the reduction in resistive power loss \cite{loss}.\\ 
Figure \ref{M2_sample} shows the temperature evolution of the \textbf{M$_2$} mode after placement of the DPPH sample. In comparison to figure \ref{M2_empty}, after placement of the sample, \textbf{f\textsubscript{res}} shifts to  lower values for the entire measurement temperature range (10 K-295 K). The dielectric property of the teflon wrapped sample perturbs the electric field (figure \ref{E_field}) on the CMSS \cite{CMSS} leading to the shift in \textbf{f\textsubscript{res}} values at different temperatures. 
The change in the transmission dip of the \textbf{M$_2$} mode as a function of externally swept \textbf{B$_0$} is recorded for obtaining the ESR spectrum at different temperature values.
The recorded ESR spectra at 10 K and room temperature along with a fit to a Lorentzian is shown in figure \ref{ESR}. 
The g-factor and ESR line width \cite{lw} are measured \& plotted for different temperature values as shown in figure \ref{result} (a) \& (b) respectively. The measured values of the g-factor and line width of the DPPH sample show satisfactory agreement with the values reported in the literature \cite{lit}.
The signal to noise ratio (SNR) \cite{SNR},\cite{SNR1} is around 46 at 10 K \& 34 at 295 K. Spin sensitivity \cite{SS},\cite{DPPH} is about 10$^{16}$ spins/gauss for the sample used with a measured ESR line width of 3.36 G at 10 K. The LSSPR based ESR spectrometer has SNR value which is one order lower than the spectrometers developed by us \cite{litho},\cite{wet} that utilize planar transmission line based resonators in reflection geometry and consequently has spin sensitivity value which is one order higher. 
\section{Conclusion}
\label{sec5}
The LSSPR has been successfully fabricated using a rapid prototyping technique, resulting in a good agreement between the measured and simulated responses. Temperature evolution of the used LSSP mode has been shown and explained. Estimated values of the g-factor and line width show good agreement with previously reported values. This is the first study to the best of our knowledge, where the magnetic field of an LSSP mode has been used to detect ESR spectra of a paramagnetic sample.\\
The simultaneous confinement of microwave electric and magnetic fields in the \textbf{M$_2$} mode can make detection of ESR signal from materials having high dielectric loss challenging \cite{loss}. However, this feature can enable one to observe
the effects of Zeeman field and Rashba field in the same ESR spectrum. The
microwave electric field induces a microwave Rashba field in
low symmetry systems (1D \& 2D samples) which can give rise
to ESR transition apart from the conventional ESR transitions caused by
microwave magnetic field \cite{CESR},\cite{CESR1}. 
\section*{Acknowledgements}
The authors gratefully acknowledge the Ministry of Education (MoE), Government of India \& Science and Engineering Research Board (SERB) (grant no.- EMR/2016/007950) for funding this work. S. R. acknowledges Council of Scientific \& Industrial Research (CSIR), India for research fellowship. The authors thank Prof. Bhaskar Gupta, Department of Electronics \& Telecommunication Engineering, Jadavpur University for providing simulation facilities. The authors are grateful to Roger Corporation, USA, for the microwave laminate samples.


\end{document}